\documentclass[12pt]{article}
\usepackage{amsfonts,amssymb,latexsym,amsthm,amscd}

\usepackage{amsmath,amsthm,amssymb}
\usepackage{graphics,graphicx}
\usepackage{epsfig}
\usepackage{graphicx}
\usepackage{epsfig}
\usepackage{epsf}
\newcommand{\half}{{\scriptstyle{\frac{1}{2}}}}
\def\lf{\left(}
\def\rg{\right)}

\begin{document}
\setlength{\baselineskip}{16pt}

\title{Correlation Energy and Entanglement Gap in Continuous Models}
\author{ L. Martina , G. Ruggeri, G. Soliani \footnote{e-mail:
          martina@le.infn.it}\\
          \mbox {
Dip. Fisica, Universit\`a del Salento, I-73100
Lecce, Italy}\\
\mbox{INFN, Sezione di Lecce, I-73100 Lecce, Italy} }

\date{}

\maketitle

\begin{abstract}
Our goal is to clarify the  relation between entanglement and
correlation energy in a bipartite system with infinite dimensional
Hilbert space. To this aim we consider the completely solvable
Moshinsky's model of two linearly coupled harmonic oscillators.
Also for small values of the couplings the entanglement of the
ground state is nonlinearly related to the correlation energy,
involving logarithmic or algebraic corrections. Then, looking for
witness observables of the entanglement, we show how to give a
physical interpretation of the correlation energy. In particular,
we have proven that there exists a set of separable states,
continuously connected  with the Hartree-Fock state, which may
have  a larger overlap with the exact ground state, but also a
larger energy expectation value. In this sense, the correlation
energy provides an entanglement gap, i.e.  an energy scale,  under
which measurements performed on the 1-particle harmonic sub-system
can discriminate  the ground state from any other separated state
of the system. However, in order to verify the generality of the
procedure, we have compared the energy distribution cumulants for
the 1-particle harmonic sub-system of the Moshinsky's model with
the case of a coupling with a damping Ohmic bath at 0 temperature.
\end{abstract}


\section{Introduction}

 The concept of entanglement has been recently considered by many authors
  in connection with several properties of the quantum systems and as a potential
   resource in quantum computation and information processing both in discrete
    and in continuous variable systems \cite{nielsen}\cite{eisert}.
    Moreover, entanglement has also been recognized to play an important role in
    the study of  many particle systems \cite{chen} and experimental
    and theoretical studies have demonstrated that it can affect
    thermodynamical properties both of the quantum phase transitions in
     the condensed matter and in molecular systems \cite{jordan}  \cite{collin}  \cite{huang}.
      However, most of the studies on the subject consider only finite dimensional Hilbert models,
       which is not the typical situation occurring in atomic/molecular
physics.  As pointed out in \cite{eisert}, the theory of the
entanglement for the infinite dimensional setting is full of
difficulties, which can be cured choosing suitable subsets of the
density matrices. In particular, the von Neumann entropy is not a
continuous function in the Hilbert space, and for any given state
of finite entanglement, one can find at least another state closer
as we want to the previous one in trace norm, which is infinitely
entangled.

However, a new area of research has been opened by \cite{huang}
\cite{dowling} \cite{mohajeri}, where it was shown that the
entanglement, even if it is not a quantum observable, can be used
in evaluating the so-called correlation energy: that is the
difference between the true eigenvalue energy of a given composite
system of identical entities, with respect to that one prescribed
by the Hartree - Fock  (HF) method. In \cite{maiolo} the case of
the formation of the Hydrogen molecule was discussed, and a
qualitative agreement between the von Neumann entropy of either
atom ( as a measure of the entanglement of formation of the whole
system) and the correlation energy as functions of the
inter-nuclear distance was shown. However, the extension of this
idea to multi-atomic molecules and its effectiveness remains still
unclear \cite{mohajeri}\cite{maiolo}. Actually, the correlation
energy is an artifact of the approximation procedure, then it is
not a physical observable and, by second, it can be modified by
the adopted method of calculations.  Nevertheless, since any other
approximating disentangled state  has also a larger energy
expectation value with respect to the HF state, one is lead to
look at the correlation energy as an entanglement gap in the sense
of \cite{dowling}. Adding any further correction term at the wave
function has to decrease the energy expectation value to the
Hamiltonian eigenvalue and increase the entanglement at the same
time up to fill the gap, describing in such a way a peculiar
domain in state space, around the exact one. Thus, the first aim
is to quantify such a kind of relation, at least on a specific
model, finding a quantitative expression of the entanglement in
terms of the correlation energy of the ground state for a
composite bipartite system. In order to have an analytically
tractable example containing all the desired features, we treated
with a family  of two coupled harmonic oscillators
\cite{moshinsky} in 3 dimensions. The coupling constant of the two
parts is in a one-to-one correspondence  with the correlation
energy and and with entanglement estimators. {\it Viceversa},
assuming that such a relation is invertible, and then an
estimation of the correlation energy can be expressed in terms of
the entanglement, one may ask how direct measurements on one of
component subsystems can provide such quantities. To this aim we
have found for the considered model an expression of the
concurrence,  in terms of the momenta of the 1-particle subsystem
energy probability distribution, from the knowledge that the
composite system is in its ground state. Thus we have an
entanglement witness and an {\it a priori} estimation of the
correlation energy. However, in concrete we have to be able  to
distinguish the energy probability distribution of the entangled
state from any other yielded by a mixed state or a pure thermal
state. First we  give an upper bound to the environmental
temperature, over which all our procedure losses validity. Then,
we compared the distribution generated by coupling a single
harmonic oscillator with an Ohmic bath at 0-Temperature, by the
analysis of all energy distribution cumulants.

In Sec. 2 we briefly review the main properties of the model: its
exact fundamental state, the HF approximation and the correlation
energy. Sec. 3  for the fundamental state of the Moshinsky's model
we evaluate the von Neumann entropy and the concurrence in terms
of the correlation energy. Since the concurrence can be expressed
in terms of the dispersions of observable conjugated quantities,
also the correlation energy takes a well defined physical meaning.
A similar relation can be established for the fidelity. In Sec. 4
we prove the existence of a continuous manifold of pure separable
states, containing the HF state, having an overlap with the exact
ground state, which can be larger with respect to the former. In
Section 5  we describe how the 1-particle energy probability
distribution for the exact state of the model can be distinguished
from that one for a single harmonic oscillator coupled with an
Ohmic bath at 0-Temperature. Some final remarks are addressed in
the Conclusions.

\section{The Moshinsky's Model}

In order to evaluate  how good the HF mean field method is in
computing quantum states,  Moshinsky  proposed a simple, but non
trivial, model of two coupled spin-$\frac{1}{2}$ harmonic
oscillators in 3 dimensions in \cite{moshinsky}. In dimensionless
unities, the Hamiltonian of the model
 reads
\begin{equation}
\hat{H}=\frac{1}{2}\left(\hat{\vec p}_1^2+\hat{\vec
p}_2^2+\hat{\vec r}_1^2+ \hat{\vec
r}_2^2\right)-\frac{1}{2}K\left(\hat{\vec r}_1- \hat{\vec
r}_2\right)^2, \label{Hamiltonian}
\end{equation}
where $\hat{\vec r}_i$ and $\hat{\vec p}_i$ denote the position
and the
 momentum operators of the i-th particle, respectively. The constant $K$
parametrizes the interaction strength of a supplementary quadratic
potential
 between the two oscillators (notice the difference of sign with respect to
\cite{moshinsky}).  The model describes a system of two identical
particles in the same harmonic potential, interacting by  a
smooth  effective repulsive coupling, which is truncated  at the
second order in a Taylor expansion for small interparticle
distances. Thus, we will dwell upon $0\leq K < 1/2$, where the
upper bound will correspond to a breaking of the model, since no
bound states can exist. This signals  that the model is far to be
realistic and it is intended  only as a toy model shaped to our
aims.

The model energy spectrum is
\begin{equation}
E_{n, m}=\frac{3}{2}\left(1+\chi^{2} \right)+n+m\chi^{2}, \quad
m,\; n \in {\bf N}\cup \left\{0\right\}
\end{equation}
where
\begin{equation}
\chi  = (1 - 2 K)^{\frac{1}{4}}, \label{chi}
\end{equation}
plays the role of effective coupling constant, parametrizing a
sort of double well potential, with an increasingly  higher (or
wider) barrier for $\chi \rightarrow 0$. The normalized position
wave function of the fundamental level is given by
\begin{equation}
\Psi _0\left({\vec  R}, {\vec  r}\right) =
\left(\frac{\chi}{\pi}\right) ^{3/2} e^{-R^2/2}e^{- \chi^2 r^2/2}
, \label{MoshWF}
\end{equation}
where the mean and relative positions
\begin{equation}
{\vec  R} = \frac{{\vec  r}_1 + {\vec  r}_2}{\sqrt{2}}, \qquad
{\vec  r} = \frac{{\vec  r}_1- {\vec  r}_2}{\sqrt{2}} \label{Var}
\end{equation}
have been defined, respectively. In general the spectrum shows
degeneracies, but the lowest level is always simple, except for
$\chi = 0$, i.e. for the limiting value of the coupling $K
\rightarrow 1/2$. Moreover, crossings occur for higher eigenvalues
at isolated points of $K$, but we are not interested to them.
Finally,  since the function (\ref{MoshWF}) is symmetric in the
interchange ${\vec r}_1\leftrightarrow {\vec r}_2$ , the total
spin must be necessarily into the singlet state. Thus, the
spinorial aspect of the problem is not relevant at this stage, and
it can be ignored.

Applying  the standard HF mean field approximation for the ground
state of the Hamiltonian (\ref{Hamiltonian}), one is led to the
wave function
\begin{equation}
\Psi _{\rm{HF}}\left({\vec  R}, {\vec  r}\right)  = \pi ^{-3/2}
(1-K)^{3/4}e^{-(1-K)^{1/2}\left.\left(R^2+r^2\right)\right/2} ,
\label{MoshHF}
\end{equation}
corresponding to the approximated eigenvalue
\begin{equation}
E_{\rm{HF}}=3(1-K)^{1/2}.
\end{equation}
Defining the {\it correlation energy} ( positive,  by Ritz's
theorem ) as
\begin{equation}
E_{\rm{corr}}  = E_{\rm{HF}} - E_{0, 0} = 3 \sqrt{1-K}-\frac{3}{2}
\left(1+\sqrt{1-2 K}\right). \label{correlation}
\end{equation}
Moreover, the explicit expression of the overlap (or the squared
fidelity) between the exact and the HF wave function  is
\begin{equation}
\left|\left\langle \Psi _{\rm{HF}}|\Psi _0\right\rangle \right|^2
=
\frac{64(1-K)^{3/2}(1-2K)^{3/4}}{\left(\left(1+\sqrt{1-K}\right)\left(1+\sqrt{1-2K}\right)-
K\right)^3} \leq 1, \label{overlap}
\end{equation}
Thus, one can figure out that adding to the HF state further
corrections,  surely  the estimation of the energy eigenvalue
improves  and the fidelity increases, but the simplest factorized
expression in (\ref{MoshHF}) will be lost.
 Differently to what happens in the approximated state,
the two oscillators in the correct fundamental state are
entangled. From analytic point of view this happens because of the
different coefficients of ${\vec R}$ and ${\vec r}$ in the wave
function (\ref{MoshWF}). From a different point of view, one can
see the expressions (\ref{MoshWF}) and (\ref{MoshHF}) as two
distinct continuous curves in the Hilbert space, parametrized by
$K$ (or $\chi$). They have only one common point at $K = 0$. The
main property of the latter curve is to contain only factorized
states.

\section{Entanglement Estimation}

Since we are dealing with pure states, the entanglement estimator
is the entanglement entropy, given in term of the  von Neumann
entropy
\begin{equation}
S_{\rm{vN}} \left[\hat{\rho }_r\right] = -\rm{Tr}\left[\hat{\rho
}_r \, \log_{ 2}\hat{\rho }_r\right] =
 - \sum _{i } \mu _{i }\, {log}_2 \left(\mu _{i }\right),
\label{eq11}
\end{equation}
of the reduced to 1-particle density matrix $\hat{\rho}_r =
\rm{Tr}_2\left[\hat{\rho }\right] =
 \rm{Tr}_1\left[\hat{\rho}\right]$,
 denoting by $\mu _{i }$ the corresponding eigenvalues.

 On the other hand, the von Neumann Entropy
$S_{\rm{vN}}$ satisfies the additive relation  $
S_{\rm{vN}}\left[\hat{ \rho } \otimes  \hat{\sigma } \right] =
S_{\rm{vN}}\left[\hat{ \rho }\right]\rm{  }+\rm{
}S_{\rm{vN}}\left[\hat{\sigma }\right], $  for any factorized
density operator $\hat{\rho }\otimes \hat{\sigma }$. But this is
precisely the structure of the reduced density matrix, which
factorizes into positional and spinorial contribution, where the
latter takes the form  $\hat{\sigma } = \frac{1}{2}\, {\bf 1} $
for the singlet spin state. Thus, it contributes to an additive
constant term (equal to 1), which  measures only the equal
uncertainty in attributing one of the two possible quantum states
to each spin. Following the ideas in \cite{ghirardi} for fermions,
 anti-symmetrizing the product of 1-particle orthogonal
 states into a spin stationary state
contains all information about entanglement  by definition.  In
conclusion, here we will compare only the contributions to the
entanglement coming from the space configurations factor of the
2-particle fundamental state.

 In the position representation the exact 2-particles density matrix $\hat{\rho }$
 for the fundamental state (\ref{MoshWF}) is given by the integral  kernel
\begin{equation}
\rho_0 \left({\vec  r }_1,{\vec  r}_2,{\vec  r '}_1,{\vec  r
'}_2\right) = \left(\frac{\chi}{\pi}\right)^{3}
e^{-\left(R^2+R'^2\right)/2}e^{-\chi^{2} \left(r^2+r'^2\right)/2},
\end{equation}
where the supplementary variables ${\vec  R'}$ and ${\vec r'}$ are
in analogy with (\ref{Var}). A similar expression holds for the
$|\Psi _{\rm{HF}}\rangle$ state (\ref{MoshHF}), where the density
matrix is given by the product of gaussian normal distributions
with the same variance. The consequences of such different
structure can be seen also by the comparison of the 1-particle
space distribution densities, which are given by
\begin{equation}
\rho _{01 }\left( {\vec  r}\right) = \left(\frac{2 \chi^2 }{\pi
\left(\chi^2 + 1\right)}\right)^{3/2} \, e^{- 2 \frac{ \chi^2
}{\chi^2 + 1}\;r^2},\; \rho _{\rm{HF1} }\left( {\vec r}\right)
=\frac{ (1-K)^{3/4}}{\pi ^{3/2}}e^{-\sqrt{1-K} \; r^2}.
\end{equation}
Thus, because of the repulsive interaction,  the exact average
distance between the particles is larger than in the approximated
estimation, being their ratio $\left(\frac{\sqrt{1-K}
\left(1+\sqrt{1-2 K}\right)}{2\sqrt{1-2 K}}\right)^{1/2}$, with a
divergence for $K\rightarrow \frac{1}{2}$ ($\chi \rightarrow 0$).

The exact 1-particle  integral operator density matrix
$\hat{\rho}_{r } = \int \rho_{r} \left({\vec r},\, {\vec
r'}\right)\; \cdot \; d\,{\vec r'}$ has the kernel
\begin{eqnarray}
\rho _{r }\left({\vec  r},{\vec  r'}\right)& = & \left(\frac{2 \chi^{2} }{ \pi \left( \chi^2 +1 \right)}\right)^{3/2}\nonumber \\
& &  \exp \left[\frac{2 \left(\chi^2 - 1\right)^2 {\vec  r} \cdot
{\vec  r}'-\left(4 \chi^2 +\left( \chi^2 + 1\right)^2\right)
\left({\vec  r}^{ 2}+\, {\vec  r}\, '^{ 2}\right)}{8 \left(\chi^2
+1\right)}\right]. \label{kernel}
\end{eqnarray}
That can be rewritten in the usual gaussian form \cite{jordan}
 \cite{EisertPlenio}
\begin{eqnarray}
\rho _{r }^{\left( \Delta p, \Delta q \right)}\left({\vec r},{\vec
r'}\right) =  \left(\frac{1}{ 2 \pi \Delta q^2}\right)^{3/2} \exp
\left[-\frac{1}{2} \left( \frac{\left({\vec r} + {\vec
r}'\right)^2}{4 \Delta q^2 }  + \Delta p^2\left({\vec r}-{\vec
r}'\right)^2 \right)\right], \label{kernelDpDq}
\end{eqnarray}
where the squared mean values for the 1-particle position and
momentum
\begin{equation}
\Delta q^2 = \frac{\chi ^2+1}{4 \chi ^2} \; , \qquad  \Delta p^2 =
\frac{1}{4} \left(\chi ^2+1\right), \label{MSVpq}
\end{equation} have been introduced.

The system becomes disentangled when the 1-particle state is the
pure minimal packet, i.e. when $\Delta p^2 \Delta q^2
=\frac{1}{4}$. But this occurs only for $K=0$ ($\chi=1$). In this
sense $\Delta p^2$ and $\Delta q^2 $ contains information about
the entanglement of the system, as remarked in \cite{jordan}.
However, differently from \cite{jordan}, by the simple algebraic
relation $\chi^2 = \frac{\Delta p^2}{ \Delta q^2} $ the squared
mean values contain also information about the form and the
strength of the interaction.  Thus, one would arise the question
if the analysis of (\ref{kernelDpDq}) not only provides
information about an entangled harmonic oscillator, but also the
main properties of the coupling: is it coupled to a small system
or to a thermodynamic bath?

From expression (\ref{kernel}) the kernel of the eigenvalue
equation for $\hat{\rho}_r$ is symmetric and of Hilbert-Schmidt
type, since the coefficient of ${\vec r}^{\;2}$ (and ${\vec r}\,
'^{\;2}$ ) is negative. So the spectrum is real and discrete, as
for the tensor product of three independent oscillators.
Accordingly, the eigenfunctions of $\hat{\rho}_r$  can be
factorized in the product of three functions, each of them
 depending only on one real variable and the eigenvalues as a product
\begin{equation}
u_{l, m, n\rm{  }}\left({\vec  r}\right) = w_l(x) w_{m }(y) w_{n
}(z), \qquad \mu _{l, m, n}= \nu _l \nu _m \nu _n , \label{18}
\end{equation}
so that the problem is reduced to solve the 1-dimensional integral
spectral problem
\begin{equation}
\int {\rho }_r (x, x')\; w_l(x')\; dx'\; =\; \nu _l \; w_l(x),
\label{IntEq1D}
\end{equation}
which has  the non degenerate spectrum and eigen-solutions of the
form
\begin{equation}
\nu _l = C c^l, \qquad w_l(x) = H_l \left( \sqrt{\chi} x \right)
\;e^{- \frac{\chi}{2}\; x^2},  \label{SolPro}
\end{equation}
where $
 C = \frac{4 \chi}{\left(1+ \chi \right)^2},\quad
  c = \left(\frac{1 - \chi }{  1  + \chi }\right)^2
$ and  $ H_l \left( \sqrt{\chi} x \right)$ denote  the Hermite
polynomials. Of course, these positive eigenvalues sum up to 1,
because they are related to a matrix density operator. On the
other hand, accordingly to (\ref{18}) the eigenvalues of the
one-particle density matrix $\hat{\rho}_r$ are given by
\begin{equation}
\mu _k= C^3 c^{ l + m + n} = C^3 c^k, \; k \,\epsilon\,
{\mathbf{N}}_{0},
\end{equation}
with degeneration  order $\deg \left[\mu _k\right]=k \left( k + 3
\right)/2 + 1$ .
 Thus, the 1-particle density
 $ \rho _{0 1 }\left({\vec  r}\right) = \sum_{l m n} C^3 c^{ l + m + n} \, | w_l\lf x \rg w_{m } \lf y \rg
 w_{n}\lf z \rg|^2 $ is represented  as a
 mixed state  in the basis of the "natural orbitals", using the
 terminology by L{\"o}dwin \cite{lowin}, which describe a  3D single harmonic oscillator of frequency
 $\chi$. The weights $\mu _k$ describe a system at the equilibrium temperature  $k_B T^* = 3 \chi/(2
 \ln\frac{1+\chi}{1-\chi})$, which is a decreasing function of
 $\chi$.
  Since experiments are always performed at a finite temperature,
  performing particle position/momentum  measurements on such a system
  we need to work at
   $T < T^* \lf \chi \rg$, in order to  highlight the quantum
   behavior discussed here.

For comparison, the spectrum of the reduced density matrix in the
the HF approximation is given by $ \left\{ 1\;\rm{(simple)}, 0
\;\rm{(infinitely \;many \; degenerate)}\right\}$, with the
corresponding eigenfunctions (for each of the three space
variables)  $ \left\{ H_l \left( \left( 1- K \right)^{1/4}  x
\right)  \exp\left[ -\frac{\left( 1- K\right)^{1/2}}{2}x^2
\right]\right\}$.

Hence, if we are allowed to look at the eigenvalues of the density
matrix operator $\hat \rho_r$  as the probabilities to find the
1-particle subsystem
 in one of the states of a ${K}$-parametrized family of harmonic oscillators,
 for small $K$ it can be found  very likely in the fundamental one.
  But this probability decreases rapidly to 0 for $ K \rightarrow \half $,
   while the higher  states become significantly more accessible.
    Notice that at $K = \half$ the system is meaningless, since all eigenvalues of
    $\hat \rho_r$ become 0 except one. However, for $0 \leq K < \half$ one can analytically sum up
    $Tr \left( \hat \rho_r \right) =1$ pointlike,
    taking into account the degeneracy. On the other hand,
    the lack of coherence can be estimated also by computing the
    $ Tr [ \hat{\rho }_r^2 ]$, which is $1$ only for  pure states.  In the present case one has explicitly
\begin{equation}
\rm{Tr}\left[\hat{\rho }_r ^2\right]= \frac{8 \chi^{3}}{\left(1+
\chi^2 \right)^3}, \label{Tracerhosquad}
\end{equation}
which is a monotonically decreasing from $1$ to $0$ function on
${ K}$.

Complementary to this quantity there is the so-called \emph{linear
entropy} \cite{buscemi},  analogous to
 the concurrence \cite{hill}
\begin{equation}
{\cal{C}}\left[\hat{\rho }_{r}\right] = 1 - \mbox{Tr}[{\hat
\rho}_r^2], \label{ConcMo}
\end{equation}
which takes values in the range $\left[ 0, 1 \right] $.
 It is invariant under local unitary transformations on the separate
 oscillators (reduced to changes of phases). This entanglement estimator seems to be quite useful
 in the present case, since  it is always bounded, even if it is defined on
 an $\infty$-dimensional Hilbert space.

The  entropy of entanglement (\ref{eq11})  can be explicitly
written as
\begin{equation}\!\!\!\!\!\!\!\!\!\!\!\!\!\!\!\!\!\!\!
 S_{\rm{vN}} \left[\hat{\rho }_{r}\right] =
 \frac{3}{\ln 4\;\chi} \left[ \left( \chi  + 1 \right)^2 \ln\left(\chi + 1\right) - 2
 \chi \,\ln 4 \chi  -\left(\chi - 1 \right)^2 \ln | \chi - 1
 |\right].
\label{MoshEntro}
\end{equation}
 For  $K \rightarrow \frac{1}{2}$ ($\chi \rightarrow 0$),
 the entropy $ S_{\rm{vN}} \left[\hat{\rho }_{r}\right]$ is logarithmic divergent,  according to the expansion
\begin{equation}
S_{\rm{vN}} \left[\hat{\rho }_{r}\right]\approx  - \frac{3 \;
\ln\left[ \chi  \right]}{\ln\left[ 2  \right]} + O\left( 1\right).
\label{approx}
\end{equation}
This is a well known result for harmonic chains
\cite{EisertPlenio}, indicating the degeneracy of the ground state
in the considered limit.

\begin{figure}
\begin{center}
\epsfig{file=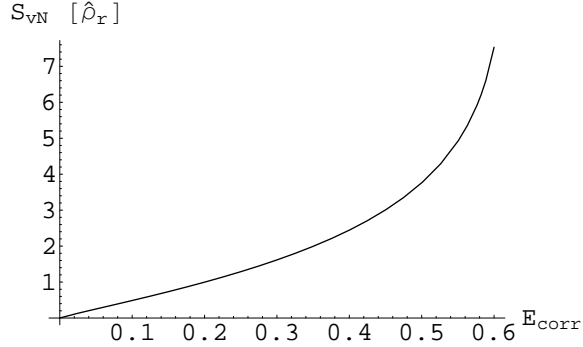, width=8cm,height=5cm} \caption{The
entanglement as a function of the correlation energy for the
Moshinsky's model. } \label{1_gr9}
\end{center}
\end{figure}

On the other hand, the  behavior of $S_{\rm{vN}}
\left[\hat{\rho}_{r}\right]$ near
 $K \rightarrow 0$ can be described by its series expansion
\begin{eqnarray}
S_{\rm vN} \left[\hat{\rho }_{r}\right] =  - \frac{3 K^2 \left(1 +
2 K \right)}{8 \, \ln (2)} \, \ln (K) + O\left(K^2\right),
\label{27}
\end{eqnarray}
approaching 0, because of  the oscillators decoupling. However,
this approximation becomes inaccurate very rapidly. From the above
expression, for $K \rightarrow 0$, the asymptotic behavior of the
entropy is controlled by a logarithmic term, differently from the
correlation energy (\ref{correlation}), which has a pure power
expansion. Then, we cannot expect a great similarity between the
two functions, also at very small values of $K$. This results
breaks the conjectured existence of a simple relationship between
the two quantities. On the other hand, let us observe that both
functions $S_{\rm{vN}} [\hat{\rho }_{r}]$ and $E_{\rm{corr}}$ are
monotonically increasing in $K$. Thus, the entanglement is an
increasing function of the correlation energy (see Fig.
\ref{1_gr9}).
 In order to have analytic expressions, we solve algebraically
the coupling constant in terms of  $E_{\rm{corr}}$ as
\begin{equation}
\chi\left( E_{\rm{corr}} \right) = \left[{1-\frac{4}{9} \left(
\sqrt{2 E_{\rm{corr}} \left(E_{\rm{corr}}+3\right) \left(2
E_{\rm{corr}}+3\right)^2}-3 E_{\rm{corr}}
\left(E_{\rm{corr}}+3\right)\right)}\right]^{1/4} \label{chiEcorr}
\end{equation}
and replacing  into $S_{\rm{vN}}[\hat{\rho }_{r}]$, we obtain a
one-to-one correspondence  $\tilde{S}_{\rm{vN}}(E_{\rm{corr}})$.

 In particular, one can look for asymptotic expressions of the entanglement for small
  values of $E_{\rm{corr}}$, corresponding to small couplings.
  Indeed, including logarithmic corrections at the lowest order near $E_{\rm{corr}} \approx  0$, one obtains
\begin{eqnarray}
\tilde{S}_{\rm{vN}}  \left(E_{\rm{corr}}\right)\approx  \frac{
\left(1+\,\ln(6)-\ln\left(E_{\rm{corr}}\right)\right)}{2 \,\ln(2)}
E_{\rm{corr}} + O\left( E_{\rm{corr}}^{3/2}\right),
\end{eqnarray}
for the Moshinsky's oscillators.  One  verifies that similar
expressions can be obtained studying other systems (for instance
the 2-points Ising model), but up to now does not exist a general
procedure to compute directly the coefficients appearing in the
above developments.   Moving to the upper limit $K \rightarrow
1/2$ ($\chi \rightarrow 0$), or equivalently  $E_{\rm{corr}}
\rightarrow \overline{E_{\rm{corr}}} = \frac{3}{2}
\left(-1+\sqrt{2}\right) $, the entropy diverges logarithmic as
\begin{equation}
\tilde{S}_{\rm{vN}}\left(E_{\rm{corr}}\right) \approx - \frac{3
\ln(\overline{E_{\rm{corr}}}-E_{\rm{corr}})}{\ln (4)}  +   O
\left( 1 \right),
\end{equation}
which is the specific behavior  for degenerate ground state, as
remarked for (\ref{approx}).

\begin{figure}
\begin{center}
\epsfig{file=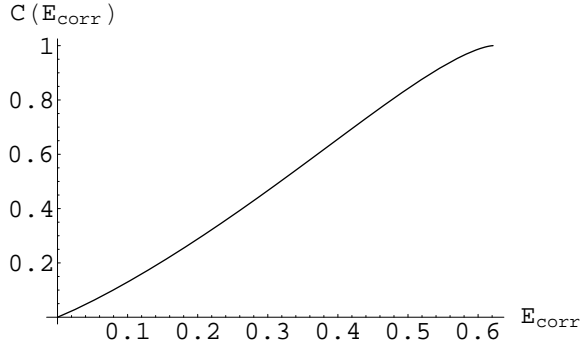, width=8cm,height=5cm} \caption{The
entanglement expressed in terms of the concurrence  as a function
of the correlation energy for the Moshinsky's model.}
\label{ConcEcorr}
\end{center}
\end{figure}

 However,  the singular behavior near 0 of the entanglement as a function of the correlation energy
 does not seem related to the specific way of its estimation. In fact,
by using Eq. (\ref{chiEcorr}) into (\ref{Tracerhosquad}), as a
function of the correlation energy the concurrence for the
Moshinsky's model  takes the form
\begin{eqnarray}
\!\! \!\! \!\! \!\!\!\! \!\! \!\! \!\! \!\!  &{\cal{C}}\left(
E_{\rm{corr}} \right)  =
  1-\frac{24 \, \sqrt{3} { {f}} ^{3/4}} {\left(f^{1/2}  + 3 \right)^3},&
\label{ConcMOEn}\\
 \quad \quad \quad\!\! \!\! \!\! \!\! \!\! \!\! \!\! \!\! \!\!  & f  =  9 +   12\, E_{\rm{corr}} \left(E_{\rm{corr}}+3\right)&-4\,
\sqrt{2} \, \sqrt{ E_{\rm{corr}} \left( E_{\rm{corr}} + 3
\right)\left(2 E_{\rm{corr}}+3\right)^2} ,\nonumber
\end{eqnarray}
a graph of which is shown in Fig. \ref{ConcEcorr}.
This function  is regular in the origin, but it is not in  its
second derivative. Again a singularity is signaling  a faster
increase of the entanglement for small values of the correlation
energy. But, the expression (\ref{ConcMOEn}) is algebraic and it
can  be manipulated more easily. Specifically, for small values of
the concurrence one gets the correlation energy   as an
half-integer power series of ${\cal{C}} $
\begin{equation}
E_{\rm{corr}} \approx {\cal{C}} + \sqrt{\frac{2}{3}}\,
{\cal{C}}^{3/2}  + \frac{2\, {\cal{C}}^2}{3} +O({\cal C}^{5/2}),
\label{EnConcMoApp}
\end{equation}
while for $ 0 \ll {\cal{C}} \leq  1$   the expansion is
\begin{equation}
E_{\rm{corr}} \approx \overline{E_{\rm{corr}}} - \frac{3}{8}
   (1-{\cal C})^{2/3} + \frac{3}{
   64} \left(\sqrt{2}-4\right)
   (1-{\cal C})^{4/3}+ O\left((1-{\cal C})\right)^2.  \label{LargeEnConcMoApp}
\end{equation}

 These expressions give  direct relations between the
correlation energy and  the entanglement, which may suggest an
experimental measure of the entanglement and of the correlation
energy. In fact, let us suppose to perform two independent series
 of measurements of position and linear momentum on one particle of the system.
Their results are  distributed with squared mean values $\Delta
q^2$ and $\Delta p^2$, respectively. On the other hand,  by
resorting to the relations (\ref{MSVpq})  in terms of the coupling
constant $\chi$ and to the expression (\ref{Tracerhosquad}), one
gets
\begin{equation} \mbox{Tr} \left[\hat{\rho }_r ^2\right] =
\left(\frac{\sqrt{\Delta p^2 \, \Delta q^2}} {\frac{1}{2}\left(
\Delta p^2 + \Delta q^2\right)}\right)^3. \label{TracerhosquadMod}
\end{equation}
Thus the entanglement is related to the ratio between the
uncertainty and the energy mean value of the observed subsystem.
Furthermore, by the definition (\ref{ConcMo}) and in the range of
validity of expansion (\ref{EnConcMoApp}) (or
(\ref{LargeEnConcMoApp}) ), one may obtain a relation among
observable quantities and the mathematical artifact
$E_{\rm{corr}}$. On the other hand, relation
(\ref{TracerhosquadMod})  has to be used carefully since, if
applied to a generic gaussian separate pure state, it does  not
give a measure of entanglement, of course. The point to be
remarked is that its value depends by the special relation of the
mean squared deviations on the coupling constant, not only on the
preparation of the state.

Finally, the fidelity of the fundamental state of the Moshinsky's
model with the corresponding HF state or, equivalently, the
overlap (\ref{overlap})   can be expressed as a function of  the
entanglement. In some sense, we are comparing two different ways
to measure the ``distance'' between the two curves of states, even
if neither quantities actually have the properties of a distance.
However, also in this case a monotonic function can be obtained
for any pair of states corresponding to the same coupling constant
$K$, or correlation energy $E_{\rm{corr}}$. This function is
regular, even if at the extrema a singularity in its higher
derivatives appears.

\section{The gap of entanglement}

 Here we would like to elucidate the special role played by  the HF state in the set of all separable
  states,  which may be closer, in the sense of the trace - norm, to the exact solution.
 To this aim and since we are looking to a neighborhood of the ground state in the Hilbert space,
  let us restrict ourselves to the pure separable states, which
 are symmetric with respect to the change $1 \leftrightarrow 2$ (the spins are into the singlet
 configuration)   and
generated by wavefunctions of the form
\begin{equation}
\tilde{\Psi} = \tilde{\Phi} \left( \vec{r}_1\right) \tilde{\Phi}
\left( \vec{r}_2 \right) , \label{factstate}
\end{equation}
where for convenience we assume that  $\tilde{\Phi}$ is normalized
to 1.
 Of course, more general choices are possible, compatibly with the assumed identity of the particles.
  In the class of states (\ref{factstate}) there exists the 1-parameter curve given by the
  gaussian functions
\begin{equation}
\tilde{\Psi}_a = \left(\frac{a}{\pi}\right)^{3/2}
\exp\left[-\frac{a}{2}\left(R^2 +r^2\right)\right],
\label{gaussfacstate}
\end{equation}
certainly containing the HF wave function (\ref{MoshHF}).
 Its overlap with the ground
state (\ref{MoshWF}) is such that
\begin{equation}
|\langle  \tilde{\Psi}_a|\Psi_0\rangle|^2 = \frac{64 a^3 \chi
^3}{(a+1)^3 \left(\chi ^2+a\right)^3}\; \geq \; |\langle
{\Psi}_{HF}|\Psi_0\rangle|^2
\end{equation}
 (see eq.  (\ref{overlap})) for $a \in\left[ a_{low} \left(\chi\right) ,
 \sqrt{\frac{1}{2} \left(\chi^4-1\right)+1} = \sqrt{1 - K}\,\right]$.
 Of course, the upper bound is exactly  the value involved in the HF wavefuction (see eq. (\ref{MoshHF})).
 The lower bound $a_{low} \left(\chi\right) $ is an algebraic positive monotonic increasing function,
 going from 0 to $\chi \rightarrow 0$, like $a_{low} \left(\chi\right) \approx
 \gamma \, \chi^2 +O\left[\chi^4 \right]$ ( $\gamma = const$) ,    to $1$ for
$\chi\rightarrow 1$ ($K \rightarrow 0$), when the two extrema
coincide.
    In particular, one sees that the maximum of overlapping is achieved for $a_{max} = \chi$,
    for which one has $|\langle  \tilde{\Psi}_{a_{max}}|\Psi_0\rangle|^2 = \frac{64\; \chi ^
 3}{(\chi +1)^6}$, equal to 1 only at $\chi=1$ ($K=0$).
  It should be noticed that $a_{max}$ is exactly the same exponent appearing in the  eigenfunctions
  of the 1-particle reduced density matrix operator (see eq.
  (\ref{SolPro})), accordingly with the notion of "natural orbital". In conclusion, the HF state is not the closest
  (in the sense of the trace-norm)
  pure separable state to the exact fundamental state and one may
  wonder if other states arbitrarily close to it may be found.
Of course, by deforming the pure gaussian form
(\ref{gaussfacstate})  with maximal overlapping, in the base of
the Hermite polynomials one can construct symmetric factorized
wavefunctions of the form
\begin{equation}
\tilde{\Psi}_{{\bf c}} = \prod_{i = 1}^3 N^{\left(i\right)}
 \left(\sum_{j = 0}^{n_i} c_j^{\left( i \right)}
 H_j \left( \sqrt{\chi} x_1^{\left(i\right)}\right)\, \right)\left(\sum_{j = 0}^{n_i} c_j^{\left( i \right)}
 H_j \left( \sqrt{\chi} x_2^{\left(i\right)}\right)  \right)\tilde{\Psi}_{a_{max}},
\label{genwavepackets}
\end{equation}
for arbitrary complex constants $\left\{c_j^{\left( i
\right)}\right\}$ and for suitable normalization constants
$N^{\left(i\right)}$. Thus, it is not difficult to find
\begin{equation}
|\langle  \tilde{\Psi}_{{\bf c}}|\Psi_0\rangle|^2 = |\langle
\tilde{\Psi}_{a_{max}}|\Psi_0\rangle|^2 \prod_{i = 1}^3
   \frac{|\sum_{j = 0}^{n_i} 2^j j!\, c_j^{\left(i\right)}{}^2 \, \left(\frac{\chi - 1}{\chi + 1}\right)^j|^2}
   {\left(\sum_{j = 0}^{n_i} 2^j j!\, |c_j^{\left(i\right)}|^2\,\right)^2}.\label{genoverlapp}
\end{equation}
The three factors in the r.h.s. of the above expression are $\leq
1$, then the overlapping of the generalized wave-packets
(\ref{genwavepackets}) with the exact state cannot exceed the
maximal one. In conclusion, we have proved that there exists a
dense set of pure and separated states, containing the HF state,
having overlap $|\langle  \Psi_{HF}|\Psi_0\rangle|^2  \leq
|\langle\tilde{\Psi}|\Psi_0\rangle|^2 < 1$, except for $\chi = 1$
($K=0$), when $\Psi_{HF} \equiv \Psi_0$. Then, the exact state
cannot be approached arbitrarily close by a separated state,
except when it is itself separate. This result is complementary to
the statement that entanglement entropy of a continuous model is a
a discontinuous function, diverging at infinity in any
neighborhood of any pure state \cite{EisertPlenio}.  The maximal
overlapping is provided by taking the suitable tensor product of
the natural orbitals \cite{lowin}. Finally, because of the
convexity of the set of all separable mixed states, i. e. of the
form $\rho = \sum_n p_n | \tilde{\Psi}_{{\bf c}_n} \rangle\langle
\tilde{\Psi}_{{\bf c}_n}|$, with $p_n \geq 0$ and $\sum_n p_n =1$,
one can extend the previous statement to the entire space of
states.

On the other hand,  the HF state has been selected in the class of
separable states by the minimal energy requirement. But in the
domain of the pure separated states of form (\ref{gaussfacstate})
the relation  $\langle \widehat{H} \rangle_{\tilde{\Psi}_a}-
E_{HF} = \frac{3 \left(a-\sqrt{1-K}\right)^2}{2 a}  \; $ holds.
Then, the expectation value of the Moshinsky's Hamiltonian
$\widehat{H}$ gets its absolute minimum indeed at the HF state.
Moreover, this can be seen also considering  general factorized
states as in (\ref{genwavepackets}). Now, because of the convexity
of the set of separable states, the minimum in the spectrum of a
bounded observable from below is always achieved  by a pure
separable state. Thus, we conclude that the above introduced
correlation energy is not simply a mathematical artifact, but it
looks analogous to the concept of {\it entanglement gap}
introduced in Ref. \cite{dowling}. Since this is a global result,
not depending on a particular computation procedure, we claim that
the HF state for the Moshinsky's  model provides the minimum
separable energy as introduced in  \cite{dowling} $E_{sep} =
\min_{\rho_{separable}} Tr \left[ \widehat{H}
\rho_{separable}\right] = E_{HF} $. Moreover, the observable
$Z_{EW} = \widehat{H} - E_{HF} {\bf 1}$ is an entanglement
witness, the spectrum of which is non negative on all separable
states and there exists the ground state (entangled) of the
Moshinsky's model for which its expectation value is $-E_{corr} <
0$. Actually, for $\chi >0$ ($K<1/2$) isolated eigenvalues of
$Z_{EW}$ may exist in the gap $\left[-E_{corr},  0 \right[$, but
they corresponds to higher energy entangled states. Thus if in a
measurement of $Z_{EW}$ we obtain a negative value, we can still
affirm that the the system is in an entangled state, even if not
necessarily in the ground state.

\section{Entanglement Energetics}
In the previous Sections we have shown that  the entanglement gap
is the main energetic scale that dictate if a composed system is,
or not, entangled. For the Moshinsky's model we have shown that
this gap is given by the correlation energy, derived from the HF
calculations. However, this relies on the full knowledge of the
density matrix, while for pure states all needed information is
encoded into the reduced matrix of a selected subsystem: in our
case one of the harmonic oscillators. Thus the question if one can
estimate the entanglement by energy measurements on the single
harmonic oscillators arises, conditionally to the knowledge that
the whole system is not in a separated ground state. These will be
subjected to statistical fluctuations, which in principle contains
the required information, i.e.  the entanglement of original
ground state of the composite system. This Section is devoted to
how extract this result and how to distinguish the energy
distribution of the entangled composite system,  from the effects
of couplings to a more generic environments, like an Ohmic bath,
even if the latter is at 0 temperature.

First step concerns the calculation of the energy distribution for
one single harmonic oscillator
 included into the Moshinsky's model. To this aim  it is useful
to have the above expressions in the simple harmonic oscillator
Hamiltonian eigenvector basis $ {\cal B} =
\left\{\varphi_{l,m,n}^i \right\}$ ( $i = 1,2 $).

First, the overlap of the exact wave function with a generic
factorized state can be evaluated from the set of amplitudes
\begin{equation}
\langle \varphi_{l,m,n}^1 \varphi_{l',m',n'}^2|\Psi_0\rangle =
\left(\frac{\chi^{1/2}}{\pi}\right)^3 \frac{\emph{I}_{l,l'}
\emph{I}_{m,m'}\emph{I}_{n,n'}}{\sqrt{2^{l+l'}2^{m+m'}2^{n+n'}l!l'!m!m'!n!n'!}},\label{reprHO}
\end{equation}
 where  the matrix $\left\{\emph{I}_{m,m'} \right\}$ has a sort
of chessboard structure, given by the relation
\begin{eqnarray} \emph{I}_{m, m'} = 2 \pi \epsilon\left(m + m'
\right)\left( -1 \right)^{m'} \left(m + m' -1 \right)!! \left(1 +
\zeta \right)^{1/2} \zeta^{\frac{m+m'}{2}}, \label{solrec}
\end{eqnarray} where the expression $\zeta = \frac{1-\chi^2}{1+\chi^2}$ and the scaled step function $\epsilon\left( m \right) = \left\{
\begin{array}{c}
1/2\;\; m\;{\rm even} \\
0 \;\; \;\;m\; {\rm odd} \\
\end{array}
\right. $ have been introduced.

On the other hand,  in the basis $ {\cal B} $  the amplitudes  of
$\tilde{\Psi}_a$ are given by
\begin{equation}
\langle\varphi_{l,m,n}^1 \varphi_{l',m',n'}^2| \tilde{\Psi}_a
\rangle =
\frac{a^{3/2}}{\pi^3}\frac{\emph{A}_{l}\emph{A}_{l'}\emph{A}_{m}\emph{A}_{m'}\emph{A}_{n}\emph{A}_{n'}}
{\sqrt{2^{l+l'}2^{m+m'}2^{n+n'}l!l'!m!m'!n!n'!}}, \label{amplHFg}
\end{equation}
where
\begin{eqnarray}
\emph{A}_{l} =
 \epsilon\left( l  \right) 2^{\,l + 1 + \frac{1}{2}} \alpha^{-l-1}
\left(2-\alpha^2\right)^{l/2} \Gamma \left(\frac{l+1}{2}\right)
\label{amplHF}
\end{eqnarray}
with $\alpha = \sqrt{1+a} $  for brevity. The expression
(\ref{amplHF}) establishes that the non vanishing terms occur only
for even principal quantum numbers $l, l', \dots $, but they are
not correlated among them.

In the representation (\ref{reprHO})-(\ref{solrec})  the elements
of the full matrix density operator for the exact ground state are
given by taking the tensor product in three dimensions of the
1-dimensional factors
\begin{eqnarray}
\rho_{l, l', m, m'}& = &4 \left(1 + \zeta \right) \chi \; \zeta^{\frac{l + l' + m + m'}{2}} \\
\nonumber  & &\frac{ \left( -1\right)^{ l' + m'} \;
\varepsilon\left( l + l'\right)
 \varepsilon\left( m + m'\right)\left( l + l' -1 \right)!!
  \left( m + m' -1 \right)!!}
  {\left[ 2^{ l + l' + m + m'} l!l'! m! m'!\right]^{1/2}}.
\end{eqnarray}
The corresponding reduced density matrix $\rho_{r, l, l'} =
\sum_{m} \rho_{l, m, l', m}$ can be computed from the above
expression, or using the continuous
 basis representation, contracting with respect the suitable states
  of the uncoupled harmonic oscillator.
  In particular, we are interested in the evaluation of the
   diagonal elements $\rho_{r, l, l}$, which represent the probabilities to find
    the 1-particle subsystem into the energy eigenstates of the uncoupled harmonic
    oscillator. It results that these quantities are related by the following recursion relation
\begin{equation}
\rho_{r, l + 1, l +1} = \frac{ l !}{\left( l + 1 \right) !}\chi \;
\left(\zeta +1 \right) \; \zeta^{l + 1}   \partial_{\zeta} \left(
\frac{\rho_{r,  l,  l}} {\chi \; \left(\zeta +1 \right) \zeta^{
l}} \right)
\end{equation}
with the  expressions for the fundamental state
\begin{equation}
 \rho_{r, 0, 0} = 2 \; \frac{\chi \; \left(\zeta +1 \right)}{\sqrt{4 - \zeta ^2}}\; .
\end{equation}
Thus, the general structure of the considered distribution is
\begin{equation}
\rho_{r, l, l} = 2 \zeta^l \frac{\chi \; \left(\zeta +1
\right)}{\left( 4 - \zeta ^2\right)^{\frac{2 l + 1}{2}}}\;
Q_l\left(\zeta\right), \label{rhoHOHO}
\end{equation}
where $Q_l \left(\zeta\right)$ is a polynomial of degree $l$ in
the "scaled" coupling constant $\zeta$. This distribution of
probability has its own peculiarities, which make it different
from a generic factorized state or
 from a pure equilibrium thermodynamical distribution.
 Then, for comparison one computes  the energy probability
 distribution  for  a factorized gaussian state by using the formulae (\ref{amplHFg}) -
 (\ref{amplHF}).
 For each eigen-state label one obtains the expression
\begin{equation}
\rho_{r, l, l} = 16 a^{1/2}\frac{ \varepsilon\left( l \right)^2
\left( l-1\right)!!^2}{l! \left( a+1\right)\left(a+3\right)^{1/2}}
\left(\frac{1-a}{1+a}\right)^l.
 \end{equation}
The first observation is that this distribution is different from
0 only for even $l$: this is a common character of all factorized
gaussian states, included the HF approximated wave-packet, so it
could be used to make an experimental comparison with the
entangled state.

\begin{figure}[p]
\centering
\includegraphics[width=47mm]{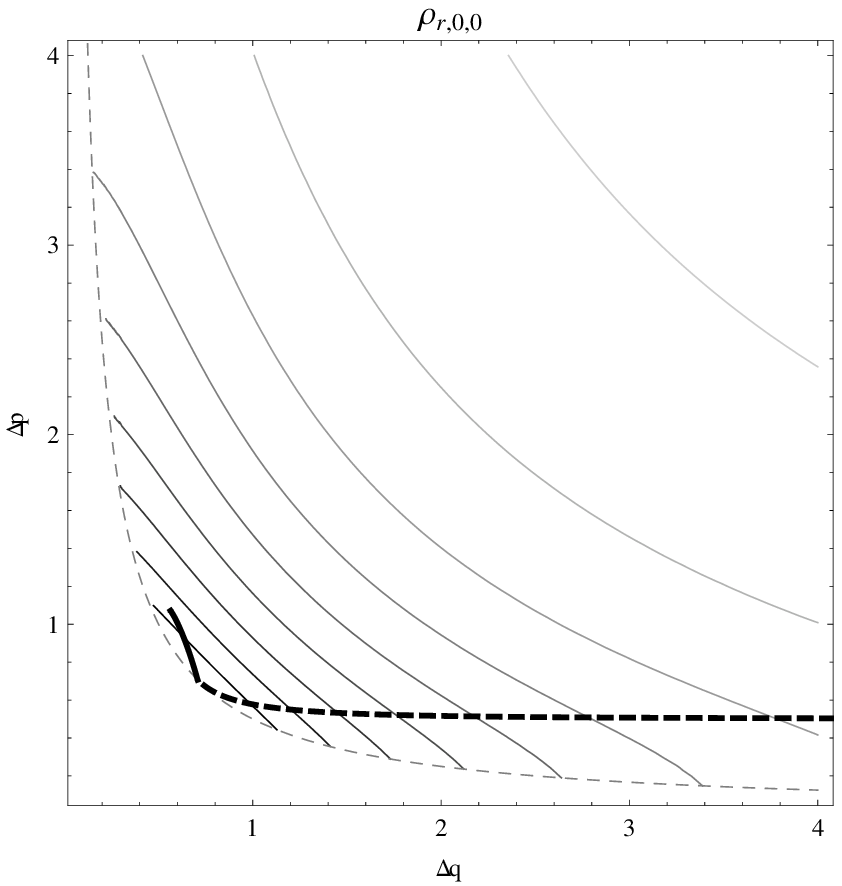}
\quad
\includegraphics[width=47mm]{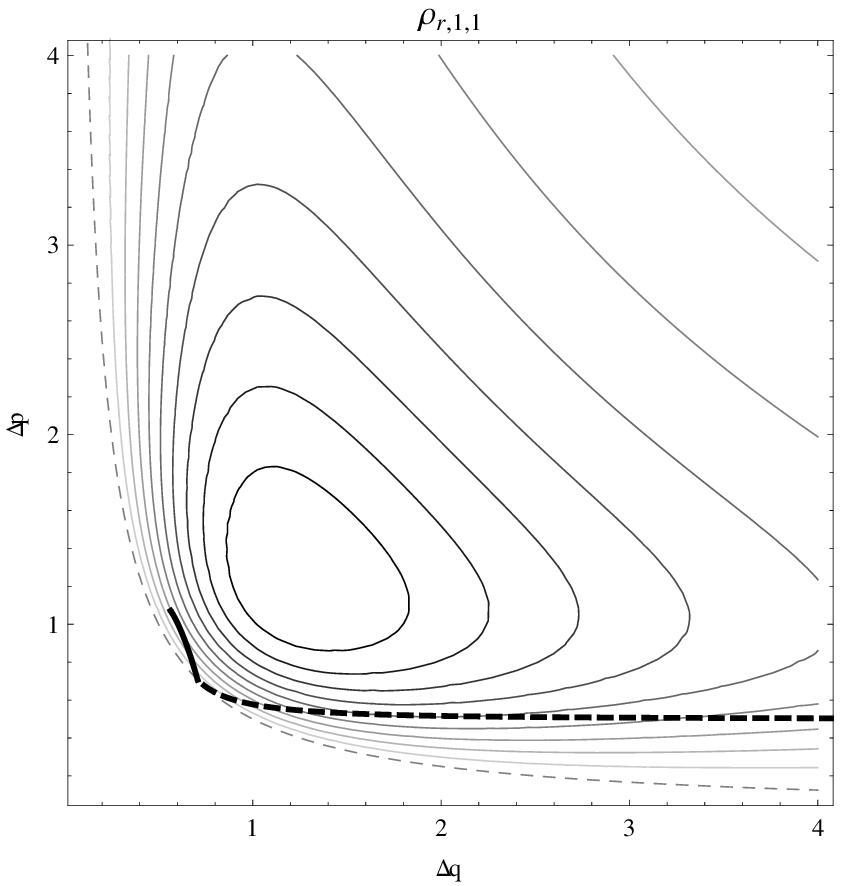}\\
\includegraphics[width=47mm]{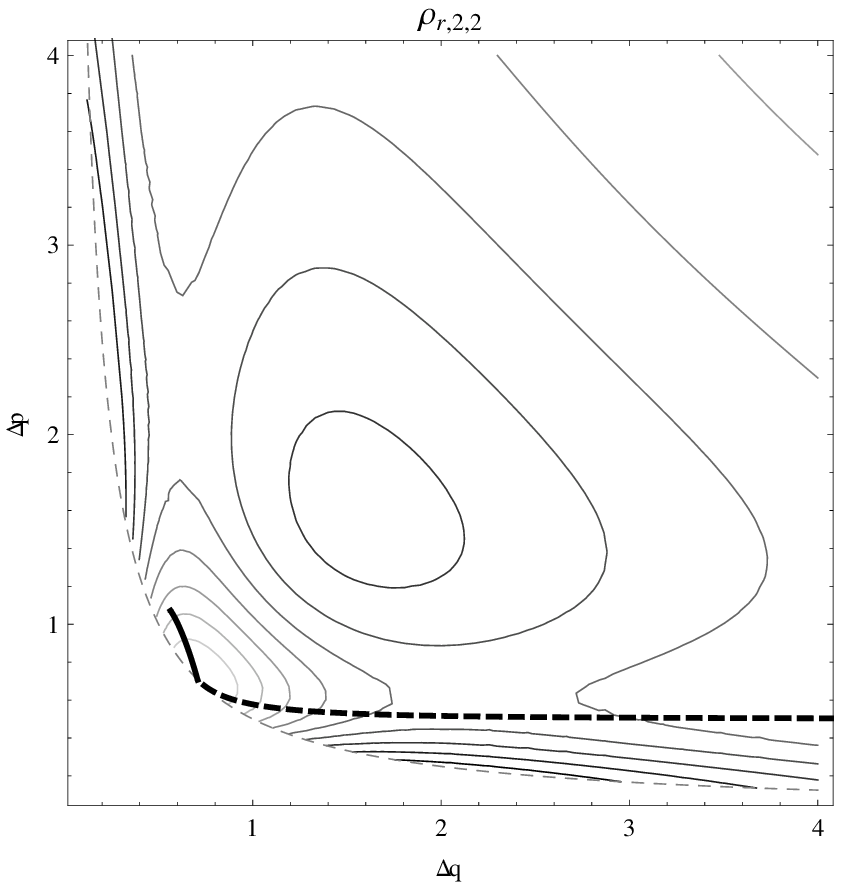} \quad
\includegraphics[width=47mm]{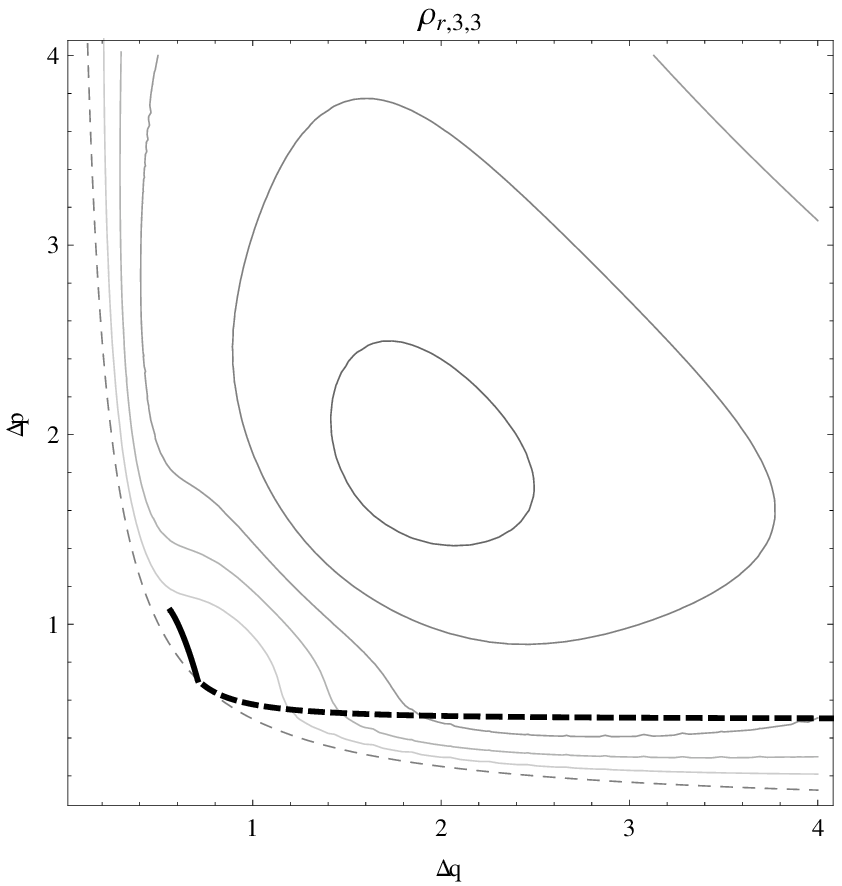} \\
\includegraphics[width=47mm]{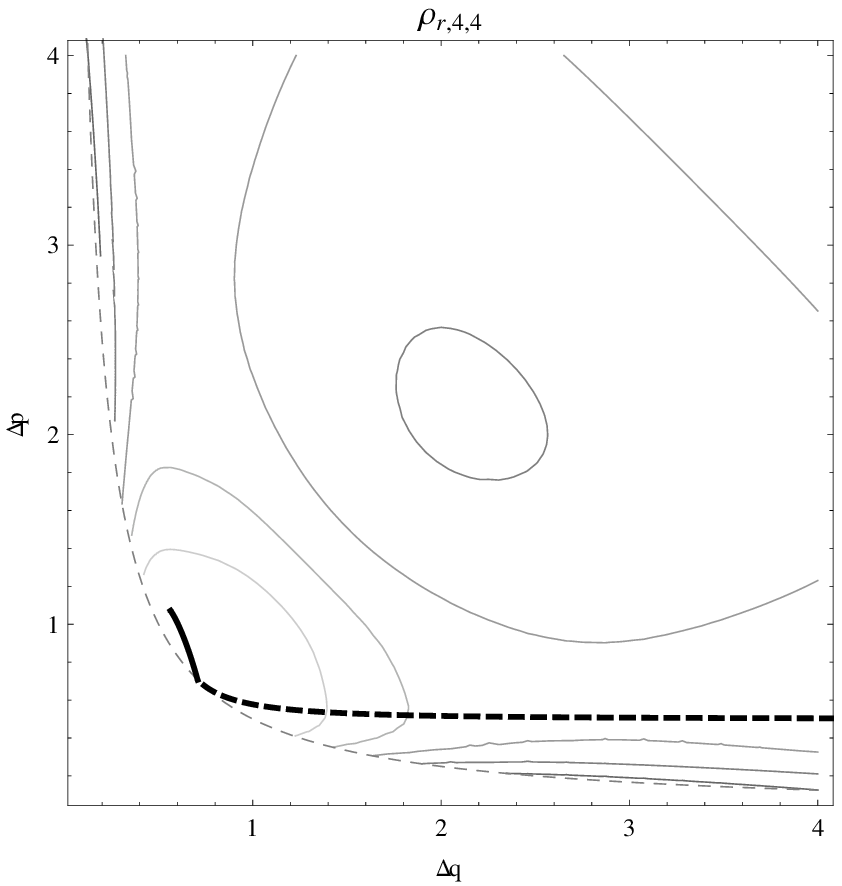}
\quad
\includegraphics[width=47mm]{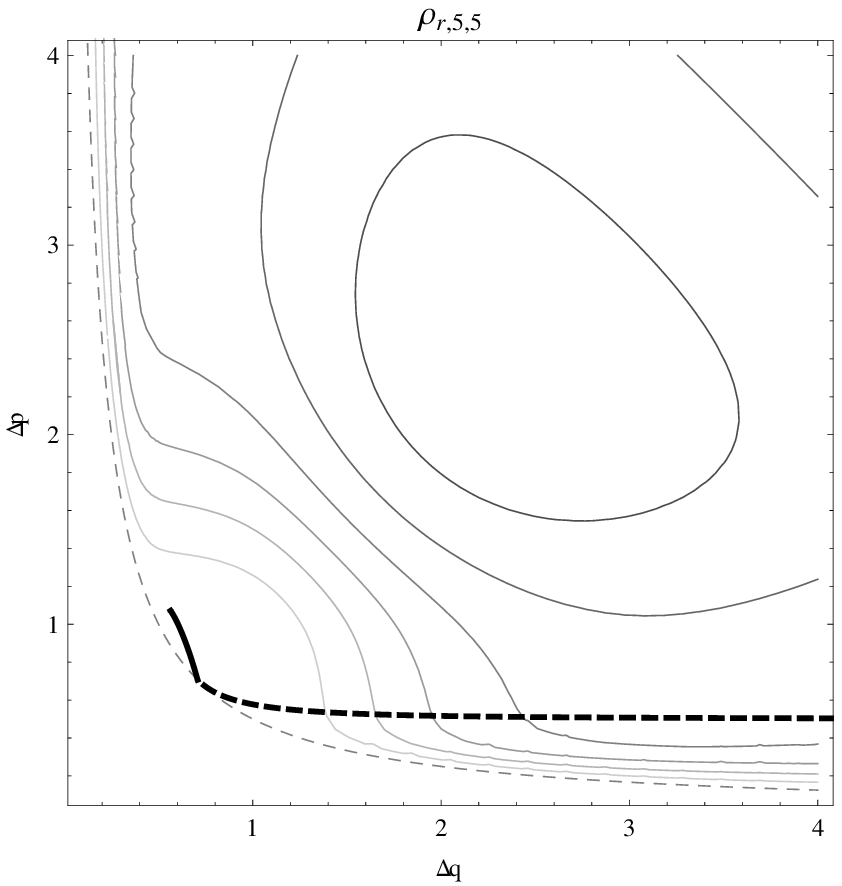}
\centering \caption{The probability distributions
(\ref{rhoDelpDelq}) to find into one of  the first six modes of
the simple harmonic oscillator a 1-particle subsystem, described
by a reduced density matrix (\ref{kernelDpDq}) in the $\left(
\Delta q, \Delta p\right)$ plane. In the area between two contours
the probability varies by  $1/10$ of the maxima values
$\{0.942809, 0.249761, 0.190042, 0.105418, 0.103428, 0.0669486\}$,
respectively for each plot, decreasing from black to lightest
gray. The black thick dashed curve represents the values of
$\left( \Delta q, \Delta p\right)$ given by (\ref{MSVpq}), while
the black thick continuous curve corresponds to the oscillator
coupled to an Ohmic bath at $T = 0^o$ accordingly to (\ref{a2}).
The dashed gray boundary curve
 is determined by the hyperbola $\Delta q
\Delta p = \frac{1}{2}$, which represents the gaussian factorized
states. When the coupling constants of both models vanishes, the
corresponding wave-packets are minimal.} \label{Six}
\end{figure}

On the other hand, one can  compute such a kind of quantity
  by the direct use of the 1-particle
reduced matrix (\ref{kernelDpDq}) \cite{jordan}. In fact, by using
the generating matrix for the Hermite polynomial, the diagonal
elements of the reduced matrix in the basis of the pure harmonic
oscillator are given by
\begin{equation}
\rho_{r, l, l}^{\left( \Delta p, \Delta q \right)} = 2 \frac{
\left[\left(2 \Delta p^2-1\right) \left(2 \Delta q^2-1\right)
\right]^{l/2} }{\left[\left(2 \Delta p^2+1\right) \left(2  \Delta
q^2+1\right)\right]^{\left( l + 1\right)/2}} P_l\left(\frac{4
\Delta p^2  \Delta q^2-1}{\sqrt{\left(4 \Delta p^4-1\right)
\left(4   \Delta q^4-1\right)}}\right), \label{rhoDelpDelq}
\end{equation}
where $P_l$ denotes the $l$-th Legendre polynomial. The parameters
$\Delta p$ and $\Delta q$ are independent quantities, limited only
by the minimal uncertainty condition $\Delta p^2 \Delta q^2 \geq
\frac{1}{4}$. Of course,  substituting the expressions of $\Delta
p$  and of $\Delta q$ given in (\ref{MSVpq}), one recovers the
formula (\ref{rhoHOHO}): there the emphasis is on the dependency
by the coupling strength.
 In Fig. (\ref{Six})
 we give a set of contour   plots of the  probabilities to find the single harmonic subsystem ( of frequency $\omega = 1$)
 in one of the first six eigenvalues
    as functions of the uncertainties $\left( \Delta p, \Delta q\right)$,
  accordingly to expression (\ref{rhoDelpDelq}).   The bold dashed
  curve is given by the equations (\ref{MSVpq}) of the uncertainties
  in the Moshinsky's model.
Of course, the efficacy of the above  procedure to measure the
entanglement has to be evaluated by comparison with other
situations. For instance, one may ask if is it possible to
distinguish the above distribution of energy eigenvalues from a
sufficiently general mixed one. Specifically we consider that one
obtained coupling one of the harmonic oscillator to an Ohmic bath
\cite{weiss}. To this aim, we propose two different methods for
this comparison.

The first way  is based on the position and momentum measurements,
from which we can reproduce the graph of Fig. \ref{Six}, for fixed
values of the coupling constant. In this case, the curves
corresponding to  the parametric representations of $(\Delta q,\,
\Delta p)$ characterize the two models. Thus, we can distinguish
the Ohmic model from the Moshinsky's model by knowing the position
and momentum uncertainty behaviour in  the $(\Delta q,\, \Delta
p)$ plane. Because of such measurements, this method produces a
lack of information about the energetics of the system. Then, the
second method, suggested by \cite{jordan}, is based on the
analysis of  the cumulants of the simple harmonic oscillator
Hamiltonian $ H_{HO}$, namely
\begin{equation}
\langle \langle H_{HO}^n \rangle \rangle = (-1)^n \left.
\frac{d^n}{d\xi^n} \ln Z(\xi) \right|_{\xi=0}, \label{cumulants}
\end{equation}
where the partition function $Z(\xi)$ is evaluated by tracing the
harmonic propagator in the imaginary time $\xi$, with respect the
generic gaussian density matrix  (\ref{kernelDpDq}). Following the
standard calculations  \cite{jordan}   \cite{Zinn}, one knows that
\begin{eqnarray}
\!\!\! \!\!\! \!\!\! \!\!\! Z(\xi) &=& \langle e^{-\xi H_{HO}}
\rangle_{\rho _{r }^{\left( \Delta p, \Delta q \right)}} = \\
\nonumber  & &\left[(\Delta p^2+\Delta q^2)\sinh \xi+2\Delta
q^2\Delta p^2(\cosh \xi-1)+\frac{1+\cosh \xi}{2}\right]^{-3/2}.
\end{eqnarray} Then, a list of the cumulants can be
algorithmically computed as   polynomials of even dergee on the
uncertainties $\left(\Delta q, \Delta p\right)$, as for instance
\begin{eqnarray}
\!\!\!\!\!\!\!\!\!\!\!\!\!\!\!\!\!\ \langle \langle H_{HO}^n
\rangle \rangle &=& \frac{3}{2}\left(\left(n-1\right)!
\left(\Delta q^{2n}+\Delta p^{2n}\right) + O\left(\Delta
q^{2\left(n-2\right)}+\Delta p^{2\left(n-2\right)}\right) \right),
\label{cumulant}
\end{eqnarray}
For the Moshinsky's model one substitutes the uncertainties given
in (\ref{MSVpq}) in (\ref{cumulant}), while for the Ohmic bath (in
the underdamped limit) one uses the mean squared values
\cite{jordan}
\begin{eqnarray}
\Delta q^2 =
\frac{1}{2\sqrt{1-\beta^2}}\left(1-\frac{2}{\pi}\arctan
\frac{\beta}{\sqrt{1-\beta^2}}\right), \nonumber \\ \Delta p^2  =
(1-2\beta^2)\Delta q^2+\frac{2\beta}{\pi}\ln\frac
{\omega_C}{\omega},\label{a2}
\end{eqnarray}
where $\beta$ is the coupling to the dissipative environment, in
units of the oscillator frequency, and $\omega_C$ is a cutoff
frequency.

\begin{figure}
\begin{center}
\epsfig{file= 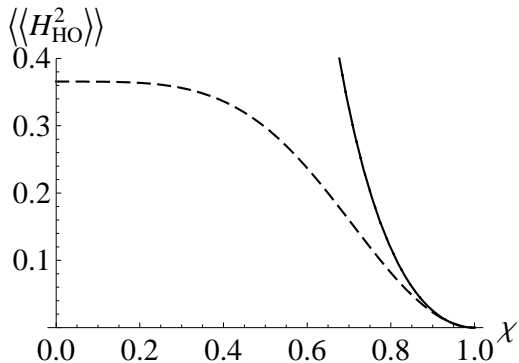,scale=.8} \caption{The second energy
cumulants $\langle\langle H_{HO}^2\rangle\rangle$
 as a function of the coupling constant
 $\chi$ with $\omega_C=10\,\omega$.
 Dashed line represents the Ohmic bath,
  while solid line corresponds to system (\ref{Hamiltonian}).}
\label{cum2}
\end{center}
\end{figure}

In order to have a unique parameter,
 which  measures the  interaction strength  between the
singled out oscillator with the remaining of the composite system
(the environment) in  both considered cases, let us  assume the
relation \begin{equation} \beta=\frac{(1-\chi^4)^2}{4(1+\chi^4)}.
\label{alpha}
\end{equation}  This relation is suggested by  the derivation of the
classical Langevin equation from a pure Hamiltonian system of
coupled oscillators  \cite{weiss}.
 Thus, we can rewrite the above cumulants in terms of the coupling constant
$\chi$ for both systems and compare them,  as  it has been shown
in Fig. \ref{cum2} for those of order 2.
There, we can see that the two functions are similar for $\chi =
1$,
 i.e. when they  describe free harmonic oscillator in both cases.
The situation  dramatically changes when $\chi$  decreases, i.e.
for stronger interactions. In fact, for $\chi\rightarrow 0$,  the
second energy cumulant associated with the Ohmic bath remains
finite, while for the Moshinsky's  model it is divergent. Similar
considerations can be made for the higher order cumulants. Thus,
we have provided a method for distinguishing the two classes of
states.

\begin{figure}
\begin{center}
\epsfig{file=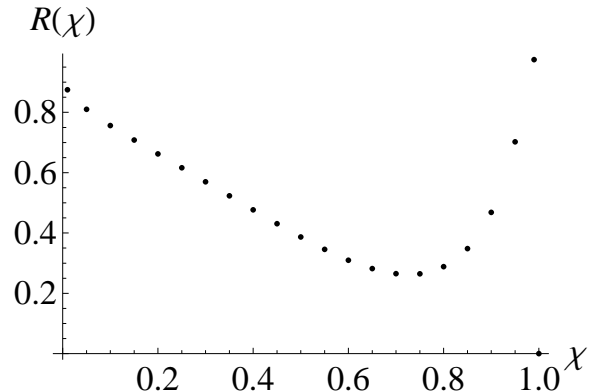,scale=.9}\caption{The relative slope
difference $R\left(\chi\right)$ as a function of the coupling
constant $\chi$.}\label{diff}
\end{center}
\end{figure}

A different approach concerns the  analysis the logarithm of the
cumulants, at a fixed value of the coupling parameter $\chi$, as
function of their order $n$.
 In fact, it results that
$\ln\langle\langle H_{HO}^n \rangle\rangle$ is approximatively a
linear function of $n$. But, from expression (\ref{cumulant}), the
relevant physical information is contained  in their slope  and in
 the corresponding  differences  between the two models.
Then, we introduce the function
\begin{equation}R(\chi) = 1- \frac{\left( \ln{ \langle \langle H_{HO-Oh}^n
\rangle\rangle}\right)_{\chi}} {\left( \ln{\langle \langle
H_{HO-Mosh}^n \rangle\rangle}\right)_{\chi}}, \label{Rchi}
\end{equation}
 which gives  the relative difference of the $\ln\langle\langle H_{HO - J}^n
  \rangle\rangle$ ($J = Oh, \; Mosh$) slopes for the two models  at different $\chi$ values (see Fig. \ref{diff}).
By inspection, we can deduce that for stronger or weaker
interactions,
 i.e. for $\chi \approx 0$ or
 $\chi \approx  1$,   the two models are well distinguishable.
While this  becomes less obvious for an intermediate range of the
coupling constant,
  where the relative  difference takes values  $R(\chi)<0.5$.

\begin{figure}
\begin{center}
\epsfig{file=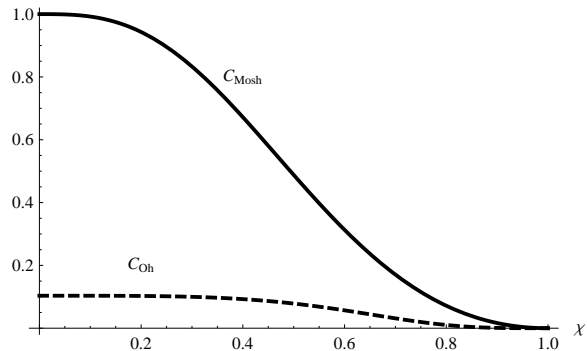,scale=.9}\caption{The concurrence for
the Moshinsky's model and the Ohmic model accordingly with the
expression (\ref{TraceEnerMom}) as a function of the coupling
constant $\chi$.}\label{ConcCompar}
\end{center}
\end{figure}

Finally, because of the explicit dependency of the cumulants from
the position/momentum uncertainties, one may express the latter in
terms of the first (mean energy) and second (variance) energy
momentum. Thus, adopting the formula (\ref{TracerhosquadMod}) as a
common expression for any model of harmonic oscillator coupled to
an environment, one provides a new expression
  of $\rm{Tr}\left[\hat{\rho }_r
^2\right]$  as
\begin{eqnarray}
  \rm{Tr}\left[\hat{\rho }_r ^2\right]  &=&   \lf \frac{\sqrt{8 \langle\langle H_{HO}^1 \rangle\rangle^2 -
  12 \langle\langle H_{HO}^2 \rangle\rangle -9}}{2
  \langle\langle H_{HO}^1 \rangle\rangle} \rg^3   \label{TraceEnerMom}
\end{eqnarray}  and then of the concurrence only in terms of
measured energy distribution properties. Moreover, from the
arguments of Section 3 we get also an {\it a priori} estimation of
the correlation energy. Now, using the  parametrization  of the
coupling constant (\ref{alpha}), one can compare the  resulting
concurrencies for the considered models (see Fig.
\ref{ConcCompar}). Qualitatively one can establish the intensity
of the entanglement generated in those different models.

\section{Conclusions}

In the present article we have  clarify the  relation between the
entanglement and correlation energy in a bipartite system with
infinite dimensional Hilbert space. We have considered the
completely solvable Moshinsky's model of two linearly coupled
harmonic oscillators, which may constitute a simple case before
studying more complicated systems, like double well potentials.
The system has a coupling constant, which can be varied in a
finite range. Thus, it continuously parametrizes two special
curves in the space states: one containing the exact ground
states, the other the separable HF states. Of course, for
vanishing coupling the two curves emerge from the same state, but
their separation can be described in terms of norm, entanglement
and energy correlation. The peculiarity of the second curve is to
lie always in a set of 0 entropy entanglement states, while along
the first one it increases monotonically (in $K$), with a
logarithmic divergence when the ground state becomes degenerated.
  On the other hand, a similar description is given in terms of
  the correlation energy, which in principle is defined only for pairs of corresponding states
  (at the same coupling constant) in the two curves. We have proved that entanglement and correlation energy are
  one-to-one along these curves, at least for the considered model. However, they are not simply proportional, but
   at small couplings they have a quite  different rate of
increasing. This phenomenon occurs not only if one uses the
entropy of entanglement for pure states, but also if one
introduces  the concurrence. However, in the considered model
certain algebraic approximated expressions of the correlation
energy in terms of the concurrence are given, so that  an artifact
of the calculation methods can take a physical  interpretation.
However, at the moment we have not a general method to compute
 directly the coefficients of such type of expansions. These could be very useful in order to have an alternative
  \textit{a priori} estimation of the errors made in numerical computations of the correct expectation values of the energy.
Such a type of relation may be useful in the studies of bipartite
systems with many inner degrees of freedom,
  like the dimers of complex molecules (see \cite{maiolo} for instance).
   In this respect
the explored concept of entanglement gap and its identification we
made with the correlation energy may play an important role: it
represents the  energy range we have to be able to measure, in
order to establish if a composed system is, or not, entangled.
However, the theory of the entanglement gap for systems with
infinite dimensional Hilbert space does not seem completely
developed as for the finite dimensional case and further
investigations are needed. In the final section we have shown
that, conditionally to the knowledge that the whole system is not
in a separated ground state, one can estimate the entanglement by
energy measurements on the single harmonic oscillators. This can
be done by two sequence of position and momentum measurements, as
well by energy measurements. The distribution of the energy
measurements is sufficiently characterized in terms of its
cumulants. This analysis enable us to compare among different
systems at 0 temperature and distinguish their ability to generate
entanglement, for instance by using the parameter $R$ introduced
in (\ref{Rchi}). Finally, via the formula (\ref{TraceEnerMom}) we
propose a new estimator of the entanglement, based on the first
two momenta of energy distribution  of the considered subsystem.
We plan to check the  how good is the present approach in
considering bipartite multi-particle systems and non linearly
coupled systems. In particular, we would like to consider
integrable systems, like in \cite{Dehesa}, in which a complete
analytic control of the calculations is at the hand.  Another
direction of research is to consider a different entropy
entanglement parameter, like the quantum version of the Tsallis
entropy \cite{Xinhua}.
\section*{Acknowledgments}
 The authors acknowledge the Italian Ministry of Scientific
Researches (MIUR) for  partial support and the INFN for  partial
support under the project Iniziativa Specifica LE41. We are
grateful to S. Pascazio for helpful discussions.

\end{document}